\documentclass[sigconf]{acmart}

\AtBeginDocument{%
  }

\setcopyright{acmlicensed}
\copyrightyear{2018}
\acmYear{2018}
\acmDOI{XXXXXXX.XXXXXXX}

\acmConference[Conference acronym 'XX]{}{June 03--05,
  2018}{Woodstock, NY}
\acmISBN{978-1-4503-XXXX-X/18/06}

\usepackage{arydshln}
\usepackage{enumitem}
\usepackage{bm}

\makeatletter
\newcommand\footnoteref[1]{\protected@xdef\@thefnmark{\ref{#1}}\@footnotemark}
\makeatother

\begin{document}

\title{A Cross-Domain Study of the Use of Persuasion Techniques in Online Disinformation}

\author{Jo\~{a}o A. Leite, Olesya Razuvayevskaya, Carolina Scarton, Kalina Bontcheva}
\affiliation{%
  \institution{University of Sheffield}
  \city{Sheffield}
  \country{UK}}
\email{{j.leite, o.razuvayevskaya, c.scarton, k.bontcheva}@sheffield.ac.uk}




\renewcommand{\shortauthors}{Trovato et al.}

\begin{abstract}
Disinformation, irrespective of domain or language, aims to deceive or manipulate public opinion, typically through employing advanced persuasion techniques. Qualitative and quantitative research on the weaponisation
 of persuasion techniques in disinformation has been mostly topic-specific (e.g., COVID-19) with limited cross-domain studies, resulting in a lack of comprehensive understanding of these strategies. This study employs a state-of-the-art persuasion technique classifier to conduct a large-scale, multi-domain analysis of the role of $16$ persuasion techniques in disinformation narratives. It shows how different persuasion techniques are employed disproportionately in different disinformation domains. We also include a detailed case study on climate change disinformation, highlighting how linguistic, psychological, and cultural factors shape the adaptation of persuasion strategies to fit unique thematic contexts.
\end{abstract}



\received{20 February 2007}
\received[revised]{12 March 2009}
\received[accepted]{5 June 2009}

\maketitle

\section{Introduction}
Disinformation campaigns, amplified by digital media, are often weaponising sophisticated \textbf{persuasion techniques} to shape public opinion by manipulating public perceptions and  advocating for certain actions \cite{jowett_propaganda_2018}.

Existing research on disinformation discourse has predominantly examined the narratives and tactics employed to influence public opinion, often touching on elements related to persuasion techniques, such as propaganda, which has a broader scope \cite{jowett_propaganda_2018}. For instance, previous studies have examined pro-Kremlin propaganda in the Russo-Ukrainian war, uncovering tactics aimed at maintaining domestic support and justifying aggression \cite{amanatullah_tell_2023}. Similarly, research into climate disinformation has highlighted that logical fallacies and conspiracy theories are used to discredit scientific consensus \cite{cook_understanding_2019}, while analyses of COVID-19 disinformation have identified narratives such as fear-mongering and conspiracy theories \cite{kotseva_trend_2023}. In natural language processing, researchers focused on datasets and models for automated persuasion technique detection \cite{da_san_martino_survey_2021, srba_survey_2024}.

While these studies provide valuable insights, they focus on individual domains, and employ different taxonomies and conceptualisations. This leads to a limited understanding of how persuasion techniques are employed by disinformation narratives in different domains. To address this gap, our study employs a state-of-the-art persuasion classifier to conduct a large-scale, cross-domain analysis of the role of persuasion techniques in disinformation. In particular, we compare the use of sixteen distinct persuasion techniques in disinformation from multiple domains (COVID-19, climate change, anti-Islam, and the Russo-Ukrainian war). This paper aims to answer the following research questions:

\begin{enumerate}[label={}, leftmargin=*, itemsep=0pt]
    \item \textbf{RQ1} Do disinformation narratives from different domains employ persuasion techniques differently? 
    \item \textbf{RQ2} Are persuasion techniques adapted linguistically, psychologically, and culturally to fit the context of domain-specific disinformation?
\end{enumerate}

By comparing statistical and qualitative patterns in the use of persuasion techniques across different domains, our study provides a deeper understanding of the ways in which disinformation seeks to manipulate and offers actionable insights for countering its effects. Our code and supplementary material are made openly available to facilitate reproducibility\footnote{\label{github}\url{https://github.com/joaoaleite/pmd}}.

\section{Methodology}




\subsection{Identifying Persuasion Techniques}
The largest dataset annotated with persuasion techniques was introduced in SemEval-2023 \cite{piskorski_semeval-2023_2023}. It consists of news articles in $9$ languages, collected from a variety of mainstream and alternative sources. It is annotated with $23$ persuasion techniques at the sentence level, with the task framed as a multi-label classification problem, allowing multiple techniques to be identified simultaneously within each sentence. In this study, we leverage the state-of-the-art persuasion classifier by \citet{razuvayevskayaComparisonParameterefficientTechniques2024}, which uses a RoBERTa-Large pretrained model fine-tuned jointly with all languages translated into English. The classifier is current ranked first on the post-competion leaderboard. 

\begin{table}[t]
\caption{Datasets domains and basic statistics.}
\label{tab:datasets}
\small
\begin{tabular}{@{}lrrrr@{}}
\toprule
\textbf{}                   & \textbf{CIDII}                                           & \textbf{COVID-19} & \textbf{\begin{tabular}[c]{@{}r@{}}Climate\\ Fever\end{tabular}} & \textbf{EUvsDisinfo}                                            \\ \midrule
\textbf{Domain}             & \begin{tabular}[c]{@{}r@{}}Islamic\\ Issues\end{tabular} & COVID-19       & \begin{tabular}[c]{@{}r@{}}Climate\\ Change\end{tabular}         & \begin{tabular}[c]{@{}r@{}}Russo-\\ Ukranian\\ War\end{tabular} \\ \hdashline
\textbf{\# Sentences}       & $1,687$                                                  & $7,369$        & $254$                                                            & $10,240$                                                        \\
\textbf{Avg. Sent. Length}  & $102.3$                                                  & $95.9$         & $109.0$                                                          & $150.7$                                                         \\
\textbf{Avg. PTs per Sent.} & $1.1$                                                    & $1.1$          & $1.2$                                                            & $1.4$                                                           \\ \bottomrule
\end{tabular}
\end{table}

To improve the reliability of our analysis, we exclude persuasion techniques with low frequency in the training set (below $2\%$). We utilise the remaining $16$ persuasion techniques which are as follows: \textit{Loaded Language} ($22\%$ of instances in the SemEval training data), \textit{Name Calling-Labeling} ($16\%$), \textit{Doubt} ($13\%$), \textit{Questioning the Reputation} ($7\%$), \textit{Exaggeration-Minimisation} ($6\%$), \textit{Appeal to Fear-Prejudice} ($5\%$), \textit{Repetition} ($3\%$), \textit{Appeal to Authority} ($3\%$), \textit{Slogans} ($3\%$), \textit{Conversation Killer} ($3\%$), \textit{Appeal to Hypocrisy} ($3\%$), \textit{Guilt by Association} ($2\%$), \textit{Appeal to Values} ($2\%$), \textit{False Dilemma-No Choice} ($2\%$), \textit{Flag Waving} ($2\%$), \textit{Causal Oversimplification} ($2\%$).

\subsection{Disinformation Domains}

We study four disinformation datasets (see Table~\ref{tab:datasets}) from diverse domains: CIDII (Islamic issues) \cite{hamed_disinformation_2023}, COVID-19 \cite{patwa_fighting_2021}, Climate Fever (climate change) \cite{diggelmann_climate-fever_2021}, and EUvsDisinfo (Russo-Ukrainian war) \cite{leite_euvsdisinfo_2024}. All texts are in English, except for EUvsDisinfo, which spans 41 additional languages where we translated the non-English texts to English using GPT4o. EUvsDisinfo also includes topics beyond the Russo-Ukrainian war, which we filtered out. Finally, only texts labelled as disinformation are analysed, with trustworthy articles excluded across all datasets. All texts are tokenised into sentences, and the persuasion classifier is used to identify persuasion techniques within each sentence.


\section{Analysis}
\begin{figure}[t]
    \centering
    \includegraphics[width=\columnwidth]{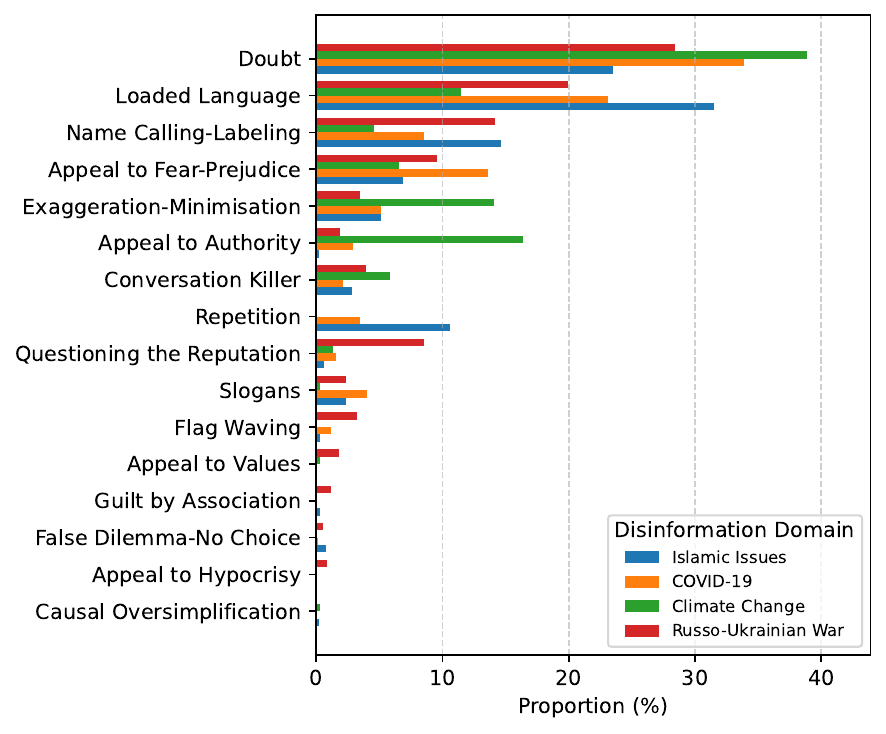}
    \caption{Proportion of persuasion techniques on the different disinformation topics.}
    \label{fig:proportions}
\end{figure}

Figure~\ref{fig:proportions} shows the proportion of persuasion techniques in each dataset, offering a visual comparison. To complement this and enable a more detailed quantitative analysis, we calculate the odds ratios (ORs) for each technique between datasets in Table~\ref{tab:odd_ratios}. We calculate ORs by comparing the odds of a technique appearing in one domain to its average proportion in the other three domains. Throughout our analysis, we only discuss statistically significant ORs, based on Fisher's exact test with $p{\mathrel{\mathord{<}}}0.05$.

\begin{table}[tb]
\caption{Odd ratios of the occurrence of each persuasion technique in one domain vs. the average proportion in the other domains. Statistically significant ratios are underlined (with $p{\mathrel{\mathord{<}}}0.05$ according to Fisher's exact test).
A$\rightarrow$Islamic issues, B$\rightarrow$COVID-19, C$\rightarrow$climate change, D$\rightarrow$Russo-Ukrainian war.}
\small
\label{tab:odd_ratios}
\begin{tabular}{lrrrr}
\toprule
\textbf{Persuasion Technique} & \textbf{A} & \textbf{B} & \textbf{C} & \textbf{D} \\
\midrule
Doubt                         & $\underline{0.60}$              & $1.18$              & $\underline{1.59}$              & $\underline{0.84}$                   \\
Loaded Language               & $\underline{2.07}$         & $\underline{1.14}$        & $\underline{0.39}$              & $0.88$                         \\
Name Calling-Labelling         & $\underline{1.72}$         & $\underline{0.74}$        & $\underline{0.34}$              & $\underline{1.61}$                   \\
Appeal to Fear-Prejudice     & $\underline{0.67}$              & $\underline{1.90}$        & $\underline{0.63}$              & $1.06$                   \\
Exaggeration-Minimisation     & $0.66$                    & $0.66$              & $\underline{3.39}$              & $\underline{0.41}$                   \\
Appeal to Authority           & $\underline{0.03}$              & $\underline{0.46}$        & $\underline{11.37}$              & $\underline{0.28}$                   \\
Conversation Killer           & $0.70$                    & $\underline{0.49}$        & \underline{$2.06$}              & $\underline{1.10}$                   \\
Repetition                    & $\underline{10.15}$        & $\underline{0.98}$        & $\underline{0.01}$              & $\underline{0.01}$                   \\
Questioning the Reputation    & $\underline{0.15}$              & $\underline{0.43}$        & $\underline{0.36}$              & $\underline{8.06}$                   \\
Slogans                       & $1.04$                    & $\underline{2.47}$        & $\underline{0.11}$              & $\underline{1.06}$                   \\
Flag Waving                   & $\underline{0.19}$              & $\underline{0.98}$        & $\underline{0.01}$              & $\underline{6.99}$                  \\
Appeal to Values              & $\underline{0.06}$              & $\underline{0.06}$        & $0.51$                    & $\underline{13.26}$                  \\
False Dilemma-No Choice       & \underline{$3.33$}              & $\underline{0.37}$        & $0.01$                    & $\underline{1.72}$                   \\
Guilt by Association          & $\underline{0.64}$              & $\underline{0.22}$        & $0.01$                    & $\underline{9.45}$                  \\
Appeal to Hypocrisy           & $\underline{0.14}$              & $\underline{0.30}$        & $0.01$                    & $\underline{19.28}$                  \\
Causal Oversimplification     & $\underline{1.97}$         & $0.13$              & $3.88$                    & $\underline{0.01}$                  \\
\bottomrule
\end{tabular}
\end{table}

We observe that \textit{Loaded Language} and \textit{Doubt} are used ubiquitously across all disinformation domains, comprising more than $20\%$ of the techniques in each, except \textit{Loaded Language} in climate change ($11\%$). Specifically \textit{Doubt} attempts to undermine trust in credible sources or established facts, creating confusion and making audiences more susceptible to accepting misleading or false claims \cite{proctor_agnotology_2008}. \textit{Loaded Language} aims to evoke strong emotional responses, such as fear or anger, which can override rational analysis and lead individuals to accept false information without critical scrutiny \cite{gennaro_emotion_2022}. 
Next, we discuss the persuasion techniques with ORs greater than $1$, with statistical significance, indicating that their usage in one domain is more prevalent than in the others.

\subsection{Domain-Specific Use of Persuasion Techniques (RQ1)}
In disinformation on Islamic issues, \textit{Name Calling-Labeling}, \textit{Causal Oversimplification}, \textit{Loaded Language}, \textit{False-Dilemma-No Choice}, and  \textit{Repetition} are the most prevalent techniques, with ORs of $1.72$, $1.97$, $2.07$, $3.33$, and $10.15$ respectively. \textit{Repetition} capitalises on the illusory truth effect, a psychological phenomenon where repeated statements are more likely to be perceived as true \cite{hasher_frequency_1977, fazio_knowledge_2015}. In this domain, \textbf{repetition is often used to associate events and/or groups}: ``\textit{All Muslim terrorists kill for the same reason the Saudi terrorists did on 9/11, the same reason ISIS killed [...], the same reason Boko Haram is killing [...], the same reason Muhammad waged wars [...]}''. \textbf{\textit{False Dilemma} forces binary choices to portray Islam as inherently extremist}, as in ``\textit{There’s no middle ground, nothing like moderate Islam}.''

In the context of COVID-19 disinformation, \textit{Loaded Language}, \textit{Appeal to Fear-Prejudice}, and \textit{Slogans} appear significantly more frequently (ORs of $1.14$, $1.90$, $2.47$, respectively). \textit{Slogans} use short and memorable phrases to convey key ideas, as in ``\textit{We need to open up, our lives depend on it}''. Both \textit{Loaded Language} and \textit{Appeal to Fear-Prejudice} are designed to evoke strong emotional responses, which can bypass rational analysis and promote the acceptance of false information \cite{gennaro_emotion_2022}. \textbf{Fear is linked to both the virus}—``\textit{\#coronavirus this is extremely scary and terrifying}''—\textbf{and vaccines}, as in ``\textit{According to Bill Gates the COVID-19 RNA vaccine will permanently alter our DNA}''. Notably, fear of COVID-19 has been linked to mental health issues during the pandemic period \cite{alimoradi_fear_2022}.

Climate change disinformation uses more frequently \textit{Doubt} ($1.59$), \textit{Conversation Killer} ($2.06$), \textit{Exaggeration-Minimisation} ($3.39$), and \textit{Appeal to Authority} ($11.37$) than the rest. \textbf{\textit{Appeal to Authority} exploits trust in credible entities like the IPCC}: ``\textit{Latest IPCC Reports show global temperature forecasts exceeded actual readings.}'' References to \textbf{NASA} and the \textbf{UN} are also common. \textbf{\textit{Exaggeration-Minimisation} downplays the urgency of climate change}: ``\textit{In the past, warming has never been a threat to life on Earth}''. \textbf{\textit{Conversation Killer} dismisses the debate with absolute certainty, leaving no room for debate}: ``\textit{Sea level rise is not going to happen.}''


Disinformation about the Russo-Ukrainian war prominently features \textit{Flag Waving} ($6.99$), \textit{Questioning the Reputation} ($8.06$), \textit{Guilt by Association} ($9.45$), and \textit{Appeal to Hypocrisy} ($19.28$). These techniques target Western and Ukrainian credibility. \textbf{\textit{Appeal to Hypocrisy} highlights perceived inconsistencies}, e.g., ``\textit{They said one thing and did another.}''—referring to NATO's territorial advancement. \textbf{\textit{Guilt by Association} links entities to controversial groups, with $28\%$ of tagged sentences referencing the word ‘Nazi’}, e.g., ``\textit{Nazi symbolism is actively utilised in their daily life}''—referring to Ukrainian soldiers. \textit{Questioning the Reputation} undermines trust in political entities: ``\textit{As for Europe, it lost its political independence after World War II}''. \textbf{\textit{Flag Waving} invokes patriotism to justify Russia's actions}, e.g., ``\textit{It is against this Evil that our soldiers bravely fight side by side}''.

\subsection{Contextual Domain Adaptation (RQ2)}

\begin{figure*}[ht]
    \centering
    \includegraphics[width=\linewidth]{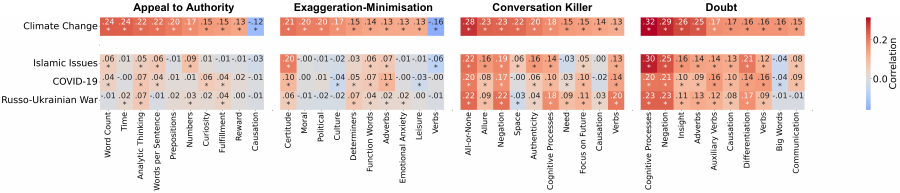}
    \caption{Top 10 most correlated LIWC features for each of the four domain-specific PTs in climate change compared to other domains. Statistically significant ($p{\mathrel{\mathord{<}}}0.05$) coefficients are indicated with an asterisk (*).}
    \label{fig:liwc}
\end{figure*}

To investigate the contextual adaptation of persuasion techniques to specific linguistic, cultural, and psychological patterns across disinformation domains, we compute the correlation between LIWC features \cite{boyd2022development} and persuasion techniques. Due to space constraints, we limit our analysis to a case study of the climate change domain, focusing on four persuasion techniques that occur disproportionately in this context: \textit{Appeal to Authority}, \textit{Conversation Killer}, \textit{Doubt}, and \textit{Exaggeration-Minimisation}. Figure~\ref{fig:liwc} presents the ten highest correlations between these techniques and the LIWC features. We compare these correlations across the other three domains to highlight patterns unique to climate change disinformation, focusing on features highly correlated within climate change but not within other domains. We provide the results for the other domains in the supplementary material\footnoteref{github}.

\textit{Appeal to Authority} in climate change disinformation uses distinct linguistic and psychological features to emphasise credibility and logic. These texts have a higher word count and words per sentence, creating longer, more complex sentences that convey authority. A high \textbf{Analytic} score reflects formal, logical language, reinforcing a well-reasoned tone. Frequent use of \textbf{prepositions} and \textbf{numbers} adds detail to enhance the legitimacy of claims. \textbf{Temporal markers} like `when' and `now' situate arguments in time, adding urgency or inevitability. Psychologically, this technique appeals to \textbf{curiosity} and \textbf{reward}, engaging readers intellectually and offering positive outcomes. Terms like `enough' and `full' (i.e., \textbf{Fulfillment}) suggest solutions framed as complete and authoritative. Notably, these texts avoid \textbf{causation} language (e.g., `because', `why'), favouring definitive statements over explanations. These patterns align with the rhetorical context of climate change narratives, where the discussion of scientific topics requires a tone of credibility and rigour to persuade audiences.

\textit{Exaggeration-Minimisation} 
employs specific linguistic and psychological features to amplify or downplay issues strategically. A high correlation with \textbf{certitude} words (e.g., `really', `of course') reinforces claims with a tone of absolute confidence, making arguments appear definitive and irrefutable. Terms related to \textbf{moral behaviour} (e.g., `honour', `deserve') inject ethical undertones, framing the issue as one of right versus wrong. Cultural references, particularly related to \textbf{politics} (e.g., `govern', `congress') and broader \textbf{culture} (e.g., `american', `chinese'), ground the discussion in societal and geopolitical contexts, often linking climate change to governance or national responsibility. Linguistically, the technique relies heavily on \textbf{determiners} (e.g., `the', `that') and \textbf{function words} (e.g., `to', `I'), creating a conversational and relatable tone. The frequent use of \textbf{adverbs} (e.g., `so', `just') adds nuance or emphasis to descriptions, subtly influencing how issues are perceived. Emotional cues, especially those tied to \textbf{anxiety} (e.g., `worry', `fear'), heighten the sense of urgency or concern, tapping into the audience's fears about climate change. Together, these features amplify certainty, invoke morality, embed discussions within cultural and political frameworks, and simplifies complex issues while evoking emotional responses.

Both \textit{Conversation Killer} and \textit{Doubt} exhibit a lesser degree of contextual adaptation, as most of the LIWC features that display high correlation in climate change are also highly correlated in other domains (e.g., Cognitive Processes, Negation). Nevertheless, some domain-specific features are present. \textit{Conversation Killer} leverages \textbf{spatial context (Space)} (e.g., ‘in’, ‘there’) to ground arguments in specific locations, \textbf{need-related states} (e.g., ‘need’, ‘have to’) to emphasise urgency and necessity, and \textbf{causation} (e.g., ‘because’, ‘why’) provides justifications that reinforce the authority of dismissive rhetoric. Similarly, \textit{Doubt} amplifies skepticism with the prominent use of \textbf{long words} to create an impression of sophistication and \textbf{communication} features (e.g., ‘say’, ‘tell’) that reference ambiguous sources, subtly eroding trust in credible information. 



\section{Discussion \& Conclusion}
In this study, we conducted a large-scale, cross-domain analysis of disinformation, examining the use and adaptation of sixteen persuasion techniques across four domains: COVID-19, climate change, Islamism, and the Russo-Ukrainian war. Our findings show that while some techniques—such as \textit{Doubt} and \textit{Loaded Language}—appear consistently across all domains, others are more context-dependent. For example, \textit{Repetition} and \textit{False Dilemma-No Choice} tend to dominate in Islamic issues, whereas \textit{Appeal to Hypocrisy} and \textit{Guilt by Association} frequently arise in the Russo-Ukrainian war domain. Beyond differences in frequency, we find that these techniques are also adapted in style. Our case study on climate change disinformation reveals that \textit{Appeal to Authority} often features more analytic and formal language than in other domains, while \textit{Exaggeration-Minimisation} frequently draws on moral and cultural references. Together, these observations demonstrate not only how disinformation messages are tailored to resonate with specific audiences, but also how their rhetorical and psychological characteristics shift according to thematic and cultural factors.

The analysis presented offers actionable insights for improving countermeasures against disinformation. Detection models can integrate domain and technique sensitive features, enhancing their ability to identify subtle, context-dependent cues. Media literacy programs can make individuals aware of the full range of persuasive strategies, including how they evolve to fit different content areas. Fact-checkers and journalists, with knowledge on how certain techniques manifest in particular domains, 
can develop targeted rebuttals that address both the presence of a technique and its adapted rhetorical form. By adopting these context-aware and targeted responses, efforts to
counter disinformation can become more effective, ultimately helping to mitigate its spread and influence.



\section*{Ethical Use of Data}
We adhere to ethical principles in data use by employing well-established and publicly available datasets that do not include personal or identifiable information. Due to the focus on disinformation topics, the data may inherently contain sensitive content.

\begin{acks}
This work has been co-funded by the UK’s innovation agency (Innovate UK) grant 10039055 (approved under the Horizon Europe Programme as vera.ai). João Leite is supported by a University of Sheffield EPSRC Doctoral Training Partnership (DTP) Scholarship.
\end{acks}

\bibliographystyle{ACM-Reference-Format}
\bibliography{bibliography}


\begin{thebibliography}{18}


\ifx \showCODEN    \undefined \def \showCODEN     #1{\unskip}     \fi
\ifx \showDOI      \undefined \def \showDOI       #1{#1}\fi
\ifx \showISBNx    \undefined \def \showISBNx     #1{\unskip}     \fi
\ifx \showISBNxiii \undefined \def \showISBNxiii  #1{\unskip}     \fi
\ifx \showISSN     \undefined \def \showISSN      #1{\unskip}     \fi
\ifx \showLCCN     \undefined \def \showLCCN      #1{\unskip}     \fi
\ifx \shownote     \undefined \def \shownote      #1{#1}          \fi
\ifx \showarticletitle \undefined \def \showarticletitle #1{#1}   \fi
\ifx \showURL      \undefined \def \showURL       {\relax}        \fi
\providecommand\bibfield[2]{#2}
\providecommand\bibinfo[2]{#2}
\providecommand\natexlab[1]{#1}
\providecommand\showeprint[2][]{arXiv:#2}

\bibitem[Alimoradi~et al.(2022)]%
        {alimoradi_fear_2022}
\bibfield{author}{\bibinfo{person}{Zainab Alimoradi~et al.}} \bibinfo{year}{2022}\natexlab{}.
\newblock \showarticletitle{Fear of {COVID}-19 and its association with mental health-related factors}.
\newblock  (\bibinfo{date}{March} \bibinfo{year}{2022}).
\newblock
\urldef\tempurl%
\url{https://doi.org/10.1192/bjo.2022.26}
\showDOI{\tempurl}


\bibitem[Amanatullah(2023)]%
        {amanatullah_tell_2023}
\bibfield{author}{\bibinfo{person}{Samy Amanatullah}.} \bibinfo{year}{2023}\natexlab{}.
\newblock \showarticletitle{Tell {Us} {How} {You} {Really} {Feel}: {Analyzing} {Pro}-{Kremlin} {Propaganda} {Devices} \& {Narratives} to {Identify} {Sentiment} {Implications}}.
\newblock  (\bibinfo{year}{2023}).
\newblock
\urldef\tempurl%
\url{https://www.illiberalism.org/tell-us-how-you-really-feel-analyzing-pro-kremlin-propaganda-devices-narratives-to-identify-sentiment-implications/}
\showURL{%
\tempurl}


\bibitem[Boyd~et al(2022)]%
        {boyd2022development}
\bibfield{author}{\bibinfo{person}{Ryan~L Boyd~et al}.} \bibinfo{year}{2022}\natexlab{}.
\newblock \showarticletitle{The development and psychometric properties of LIWC-22}.
\newblock \bibinfo{journal}{\emph{Austin, TX: University of Texas at Austin}} (\bibinfo{year}{2022}).
\newblock
\urldef\tempurl%
\url{https://doi.org/10.13140/RG.2.2.23890.43205}
\showDOI{\tempurl}


\bibitem[Cook(2019)]%
        {cook_understanding_2019}
\bibfield{author}{\bibinfo{person}{John Cook}.} \bibinfo{year}{2019}\natexlab{}.
\newblock \showarticletitle{Understanding and {Countering} {Misinformation} {About} {Climate} {Change}}.
\newblock
\showISBNx{978-1-5225-8535-0}
\urldef\tempurl%
\url{https://doi.org/10.4018/978-1-5225-8535-0.ch016}
\showDOI{\tempurl}


\bibitem[Da~San Martino~et al.(2020)]%
        {da_san_martino_survey_2021}
\bibfield{author}{\bibinfo{person}{Giovanni Da~San Martino~et al.}} \bibinfo{year}{2020}\natexlab{}.
\newblock \showarticletitle{A Survey on Computational Propaganda Detection}.
\newblock
\urldef\tempurl%
\url{https://doi.org/10.24963/ijcai.2020/672}
\showDOI{\tempurl}


\bibitem[Diggelmann~et al.(2021)]%
        {diggelmann_climate-fever_2021}
\bibfield{author}{\bibinfo{person}{Thomas Diggelmann~et al.}} \bibinfo{year}{2021}\natexlab{}.
\newblock \bibinfo{title}{{CLIMATE}-{FEVER}: {A} {Dataset} for {Verification} of {Real}-{World} {Climate} {Claims}}.
\newblock
\newblock
\urldef\tempurl%
\url{https://doi.org/10.48550/arXiv.2012.00614}
\showDOI{\tempurl}


\bibitem[Fazio~et al.(2015)]%
        {fazio_knowledge_2015}
\bibfield{author}{\bibinfo{person}{Lisa~K. Fazio~et al.}} \bibinfo{year}{2015}\natexlab{}.
\newblock \showarticletitle{Knowledge does not protect against illusory truth}.
\newblock  (\bibinfo{date}{Oct.} \bibinfo{year}{2015}).
\newblock
\showISSN{1939-2222}
\urldef\tempurl%
\url{https://doi.org/10.1037/xge0000098}
\showDOI{\tempurl}


\bibitem[Gennaro and Ash(2022)]%
        {gennaro_emotion_2022}
\bibfield{author}{\bibinfo{person}{Gloria Gennaro} {and} \bibinfo{person}{Elliott Ash}.} \bibinfo{year}{2022}\natexlab{}.
\newblock \showarticletitle{Emotion and {Reason} in {Political} {Language}}.
\newblock  \bibinfo{volume}{132}, \bibinfo{number}{643} (\bibinfo{date}{April} \bibinfo{year}{2022}).
\newblock
\showISSN{0013-0133}
\urldef\tempurl%
\url{https://doi.org/10.1093/ej/ueab104}
\showDOI{\tempurl}


\bibitem[Hamed et~al\mbox{.}(2023)]%
        {hamed_disinformation_2023}
\bibfield{author}{\bibinfo{person}{Suhaib~Kh Hamed}, \bibinfo{person}{Mohd Juzaiddin~Ab Aziz}, {and} \bibinfo{person}{Mohd~Ridzwan Yaakub}.} \bibinfo{year}{2023}\natexlab{}.
\newblock \showarticletitle{{Disinformation} {Detection} {About} {Islamic} {Issues} {on} {Social} {Media} {Using} {Deep} {Learning} {Techniques}}.
\newblock  (\bibinfo{date}{July} \bibinfo{year}{2023}).
\newblock
\showISSN{0127-9084}
\urldef\tempurl%
\url{https://doi.org/10.22452/mjcs.vol36no3.3}
\showDOI{\tempurl}


\bibitem[Hasher et~al\mbox{.}(1977)]%
        {hasher_frequency_1977}
\bibfield{author}{\bibinfo{person}{Lynn Hasher}, \bibinfo{person}{David Goldstein}, {and} \bibinfo{person}{Thomas Toppino}.} \bibinfo{year}{1977}\natexlab{}.
\newblock \showarticletitle{Frequency and the conference of referential validity}.
\newblock  (\bibinfo{date}{Feb.} \bibinfo{year}{1977}).
\newblock
\showISSN{0022-5371}
\urldef\tempurl%
\url{https://doi.org/10.1016/S0022-5371(77)80012-1}
\showDOI{\tempurl}


\bibitem[Jowett and O'Donnell(2018)]%
        {jowett_propaganda_2018}
\bibfield{author}{\bibinfo{person}{Garth~S. Jowett} {and} \bibinfo{person}{Victoria O'Donnell}.} \bibinfo{year}{2018}\natexlab{}.
\newblock \bibinfo{booktitle}{\emph{Propaganda \& {Persuasion}}}.
\newblock
\showISBNx{978-1-5063-7135-1}
\urldef\tempurl%
\url{https://books.google.com/books?id=vIC92PdJ0l0C}
\showURL{%
\tempurl}


\bibitem[Kotseva~et al.(2023)]%
        {kotseva_trend_2023}
\bibfield{author}{\bibinfo{person}{Bonka Kotseva~et al.}} \bibinfo{year}{2023}\natexlab{}.
\newblock \showarticletitle{Trend analysis of {COVID}-19 mis/disinformation narratives–{A} 3-year study}.
\newblock  (\bibinfo{date}{Nov.} \bibinfo{year}{2023}).
\newblock
\showISSN{1932-6203}
\urldef\tempurl%
\url{https://doi.org/10.1371/journal.pone.0291423}
\showDOI{\tempurl}


\bibitem[Leite~et al.(2024)]%
        {leite_euvsdisinfo_2024}
\bibfield{author}{\bibinfo{person}{João~A. Leite~et al.}} \bibinfo{year}{2024}\natexlab{}.
\newblock \showarticletitle{{EUvsDisinfo}: {A} {Dataset} for {Multilingual} {Detection} of {Pro}-{Kremlin} {Disinformation} in {News} {Articles}}.
\newblock
\showISBNx{9798400704369}
\urldef\tempurl%
\url{https://doi.org/10.1145/3627673.3679167}
\showDOI{\tempurl}


\bibitem[Patwa~et al.(2021)]%
        {patwa_fighting_2021}
\bibfield{author}{\bibinfo{person}{Parth Patwa~et al.}} \bibinfo{year}{2021}\natexlab{}.
\newblock \showarticletitle{Fighting an {Infodemic}: {COVID}-19 {Fake} {News} {Dataset}}.
\newblock
\showISBNx{978-3-030-73696-5}
\urldef\tempurl%
\url{https://doi.org/10.1007/978-3-030-73696-5_3}
\showDOI{\tempurl}


\bibitem[Piskorski~et al.(2023)]%
        {piskorski_semeval-2023_2023}
\bibfield{author}{\bibinfo{person}{Jakub Piskorski~et al.}} \bibinfo{year}{2023}\natexlab{}.
\newblock \showarticletitle{{SemEval}-2023 {Task} 3: {Detecting} the {Category}, the {Framing}, and the {Persuasion} {Techniques} in {Online} {News} in a {Multi}-lingual {Setup}}.
\newblock
\urldef\tempurl%
\url{https://doi.org/10.18653/v1/2023.semeval-1.317}
\showDOI{\tempurl}


\bibitem[Proctor and Londa(2008)]%
        {proctor_agnotology_2008}
\bibfield{author}{\bibinfo{person}{Robert~N Proctor} {and} \bibinfo{person}{Schiebinger Londa}.} \bibinfo{year}{2008}\natexlab{}.
\newblock \bibinfo{booktitle}{\emph{Agnotology: {The} making and unmaking of ignorance}}.
\newblock
\urldef\tempurl%
\url{https://philarchive.org/archive/PROATM}
\showURL{%
\tempurl}


\bibitem[Razuvayevskaya~et al.(2024)]%
        {razuvayevskayaComparisonParameterefficientTechniques2024}
\bibfield{author}{\bibinfo{person}{Olesya Razuvayevskaya~et al.}} \bibinfo{year}{2024}\natexlab{}.
\newblock \showarticletitle{Comparison between Parameter-Efficient Techniques and Full Fine-Tuning}.
\newblock  (\bibinfo{date}{May} \bibinfo{year}{2024}).
\newblock
\showISSN{1932-6203}
\urldef\tempurl%
\url{https://doi.org/10.1371/journal.pone.0301738}
\showDOI{\tempurl}


\bibitem[Srba~et al.(2024)]%
        {srba_survey_2024}
\bibfield{author}{\bibinfo{person}{Ivan Srba~et al.}} \bibinfo{year}{2024}\natexlab{}.
\newblock \bibinfo{title}{A {Survey} on {Automatic} {Credibility} {Assessment} of {Textual} {Credibility} {Signals} in the {Era} of {Large} {Language} {Models}}.
\newblock
\newblock
\urldef\tempurl%
\url{https://doi.org/10.48550/arXiv.2410.21360}
\showDOI{\tempurl}


\end{thebibliography}


\end{document}